\documentclass{aa}  
\usepackage{graphicx}
\usepackage{epsfig}
\usepackage{txfonts}
\def\spose#1{\hbox to 0pt{#1\hss}}
\def\simlt{\mathrel{\spose{\lower 3pt\hbox{$\mathchar"218$}}
     \raise 2.0pt\hbox{$\mathchar"13C$}}}
\def\simgt{\mathrel{\spose{\lower 3pt\hbox{$\mathchar"218$}}
     \raise 2.0pt\hbox{$\mathchar"13E$}}}

\begin{document}

	\title{The effect of differential galactic winds on the chemical 
evolution of galaxies}
        \author{Simone Recchi\inst{1}\thanks{recchi@oats.inaf.it} 
        \and Emanuele Spitoni\inst{2}\thanks{spitoni@oats.inaf.it}
        \and Francesca Matteucci\inst{1, 2}\thanks{matteucc@oats.inaf.it}
        \and Gustavo A. Lanfranchi\inst{3}
\thanks{gustavo.lanfranchi@unicsul.br}}
	\offprints{S. Recchi}
	\institute{
	INAF - Osservatorio Astronomico di Trieste, Via G.B. Tiepolo 11, 
		34131, Trieste, Italy \and
	Department of Astronomy, Trieste University, Via G.B. Tiepolo 11, 
		34131, Trieste, Italy \and
        N\'ucleo de Astrof\'isica Te\'orica, Universidade Cruzeiro do Sul,
                R. Galv\~ao Bueno 868, Liberdade, 01506-000, S\~ao Paulo, 
                SP, Brazil
}
	\date{Received  /  Accepted   }
	\abstract
  {}
  {The aim of this paper is to study the basic equations of the
  chemical evolution of galaxies with gas flows.  In particular, we
  focus on models in which the outflow is differential, namely in
  which the heavy elements (or some of the heavy elements) can leave
  the parent galaxy more easily than other chemical species such as H
  and He.}
  {We study the chemical evolution of galaxies in the framework of
  simple models, namely we make simplifying assumptions about the
  lifetimes of stars and the mixing of freshly produced metals.  This
  allows us to solve analytically the equations for the evolution of
  gas masses and metallicities.  In particular, we find new analytical
  solutions for various cases in which the effects of winds and infall are 
  taken into account.}
  {Differential galactic winds, namely winds carrying out
  preferentially metals, have the effect of reducing the global
  metallicity of a galaxy, with the amount of reduction increasing
  with the ejection efficiency of the metals.  Abundance ratios are
  predicted to remain constant throughout the whole evolution of the
  galaxy, even in the presence of differential winds.  One way to
  change them is by assuming differential winds with different
  ejection efficiencies for different elements.  However, simple
  models apply only to elements produced on short timescales, namely
  all by Type II SNe, and therefore large differences in the ejection
  efficiencies of different metals are unlikely.}
  {Variations in abundance ratios such as [O/Fe] in galaxies, without
  including the Fe production by Type Ia supernovae, can in principle
  be obtained by assuming an unlikely different efficiency in the loss
  of O relative to Fe from Type II supernovae.  Therefore, we conclude
  that it is not realistic to ignore Type Ia supernovae and that the
  delayed production of some chemical elements relative to others
  (time-delay model) remains the most plausible explanation for the
  evolution of $\alpha$-elements relative to Fe.}


\titlerunning{Differential winds and chemical evolution of galaxies}
\maketitle

\section{Introduction}
\label{intro}

When a stellar system is formed out of gas, the dying stars enrich the
remaining gas so that new stars are metal-enriched compared to their
ancestors.  Simple models of chemical evolution have been used since
decades to interpret the evolution of the chemical composition of
stars and gas.  Pioneering works have determined the evolution of the
metallicity in the Galaxy (Schmidt \cite{sch63}; Searle \& Sargent
\cite{ss72}; Tinsley \cite{tin74}; Pagel \& Patchett \cite{pp75}) by
making very simplifying hypotheses about the initial mass function
(IMF), the lifetime of stars and the mixing of chemical elements with
the surrounding interstellar medium (ISM).

These models remain useful guides for understanding the main chemical
properties of galaxies, but it was soon realized that they cannot give
a complete picture of galaxy evolution.  In particular, simple closed
box models cannot reproduce the distribution of metallicity of
long-living stars in the galaxy ({\it G-dwarf problem}, Lynden-Bell
\cite{lyn75}; Rana \cite{rana91} and references therein).  The most
obvious solution to the G-dwarf problem is to relax the hypothesis
that the solar neighborhood evolved as a closed box and to allow for
infall of gas (Tinsley \cite{tin80}).  Increasingly complex models of
chemical evolution including gas flows (infall, outflow, radial
flows), but always assuming the instantaneous recycling approximation
(IRA)\footnote{IRA assumes that stars with masses larger than 1
M$_{\odot}$ die instantaneously whereas those below 1 M$_{\odot}$ live
forever (Matteucci \cite{mat01})}, have been proposed, covering a large
variety of galactic morphologies (Larson \cite{lar76}; Clayton
\cite{clay88}; Edmunds \cite{edm90}; K\"oppen \cite{koe94}; Martinelli
\& Matteucci \cite{mm00}).  These models consider the global
metallicity Z and neglect metallicity-enriched outflows, so they
cannot be used to predict variations of abundance ratios.  Pagel
\& Tautvai\v{s}ien\.{e} (\cite{pt98}) have calculated abundance ratios
using analytical models, but they had to assume that chemical elements
such as iron are ejected with a fixed time delay $\Delta$ after the
time of star formation.  This is a rough approximation since it is
known that the progenitors of SNeIa span a very wide lifetime (see
e.g. Matteucci \& Recchi \cite{mr01}).  Only detailed numerical models
relaxing IRA and including the chemical enrichment from Type Ia SNe
(e.g. Matteucci \& Greggio \cite{mg86}) allow one to follow in detail
the evolution of single elements.  Matteucci \& Greggio (\cite{mg86})
showed in detail the effect of the time-delay model, already suggested
by Tinsley (\cite{tin80}) and Greggio \& Renzini (\cite{gr83}), in
particular the effect of a delayed Fe production by Type Ia SNe on
abundance ratios involving $\alpha$-elements (O, Mg, Si).

Outflows and galactic winds are produced when the thermal energy of a
galaxy (produced by energetic events such as SN explosions and stellar
winds) is enough to unbind a large fraction of the ISM (Mathews \&
Baker \cite{mb71}).  The existence of {\it differential winds}, namely
galactic winds in which the metals (or some of them) are ejected out
of the parent galaxy more efficiently than the other elements, has
been first hypothesized and applied to the chemical evolution of
galaxies by Pilyugin (\cite{pil93}) and Marconi et al. (\cite{mmt94}).
In particular, Marconi et al. (\cite{mmt94}) suggested that the
chemical properties of gas-rich dwarf irregular galaxies, most
remarkably the spread in H/He vs. O/H and N/O vs. O/H, can be more
easily explained by assuming differential winds, in which products of
SNeII (mostly $\alpha$-elements and one third of iron) can escape the
potential well of a galaxy more easily than other elements formed in
low and intermediate mass stars.  Hereafter we will call {\it
selective winds} the differential winds in which different metals have
different ejection efficiencies.  The existence of differential and
selective winds has been theoretically confirmed by several detailed
chemodynamical simulations (MacLow \& Ferrara \cite{mf99}; D'Ercole \&
Brighenti \cite{db99}; Recchi et al. \cite{rmd01}; Fujita et
al. \cite{fuj04}) and it has been observationally confirmed by
determining the metallicity of outflows in starburst galaxies (Martin
et al. \cite{mkh02}; Ott et al. \cite{owb05}).  Recently, Salvadori et
al. (\cite{sfs08}) have suggested differential winds coupled with
relaxation of IRA in massive stars to explain the steep decline of the
[O/Fe] ratio observed in the stars of Sculptor, without invoking any Fe
production from Type Ia SNe.

In this paper we apply the hypothesis of differential and selective
winds in the framework of simple models of chemical evolution with
IRA.  The chemical evolution of galaxies is a complex phenomenon that
can be properly treated only by means of detailed numerical codes,
possibly taking into account also the hydrodynamical evolution of the
system, as we have explained above.  However, in spite of the
oversimplifications introduced by simple chemical evolution models,
the results they produce are very linear and simple to interpret.
They can therefore illustrate in a direct way the effect of
differential winds on the chemical evolution of galaxies, as well as
the effect of relaxing IRA.  Moreover, for clarity purposes, we show
also the results of detailed numerical models (Lanfranchi \& Matteucci
\cite{lm03}; \cite{lm04}) as a comparison.

The plan of the paper is as follows: in Sect. 2 we summarize the
assumptions of the simple models of chemical evolution and the results
of models including gas flows (infall and outflow).  In Sect. 3 we
introduce the assumptions necessary to solve the equations of chemical
evolution of galaxies with differential winds and produce some
illustrative result.  In Sect. 4 we study the abundance ratios of some
chemical elements predicted by models with differential and selective
winds.  Finally, we discuss some implications of the results and draw
the main conclusions in Sect. 5.

\section{Summary of simple model containing 
gas flows}
\label{simple}

\subsection{Standard solutions}
\label{sec:standard}

As it is well known, the so-called {\it Simple Model} of chemical
evolution is based on the following assumptions:

\begin{enumerate}

\item the system is one-zone and closed, namely there are no inflows 
or outflows.

\item The initial gas is primordial (no metals).

\item The IMF is constant in time.

\item The gas is well mixed at any time ({\it instantaneous mixing}).

\item Stars more massive than 1 M$_\odot$ die instantaneously; stars 
smaller than 1 M$_\odot$ live forever ({\it instantaneous recycling
approximation} or IRA).

\end{enumerate}
\noindent
These simplifying assumptions allow us to calculate analytically the
chemical evolution of the galaxies.  Once we have defined the
fundamental quantities:

\begin{equation}
R = \int_1^\infty (m - M_R) \phi (m) dm,
\label{eq:r}
\end{equation}
\noindent
(where $\phi (m)$ is the IMF and $M_R$ is the mass of the remnant) and

\begin{equation}
y_Z = {1 \over {1 - R}} \int_1^\infty m p_{Z, m} \phi (m) dm,
\label{eq:yield}
\end{equation}
\noindent
(where $p_{Z, m}$ is the fraction of newly produced and ejected metals
by a star of mass $m$), the well known solution of the closed box
model can be easily found:

\begin{equation}
Z = y_Z \ln (\mu^{-1}),
\label{eq:simple}
\end{equation}
\noindent
where $\mu$ is the gas fraction $M_g/M_{tot}$.  This result is
obtained by assuming that the galaxy initially contains only gas and
has the remarkable property that it does not depend on the particular
star formation history the galaxy experiences.  To obtain eq.
(\ref{eq:simple}) one has to assume that $y_Z$ does not depend on
metallicity.  Nucleosynthesis can indeed depend on metallicity,
especially for secondary elements.  However, for O and Fe the
metallicity effect is not so important (Matteucci \cite{mat01};
Garnett \cite{garn02}), therefore we can neglect it for the purposes
of our work.  In the general case in which $y_Z = f (Z)$ the evolution
of the metallicity is governed by the following equation:

\begin{equation}
\int_0^Z {dw \over f (w)} = \ln (\mu^{-1}).
\end{equation}

Relaxing the first of the assumptions of the simple model, we get the
models including gas flows, also known as {\it leaky box} models.
Analytical solutions of simple models of chemical evolution including
infall or outflow are known since at least 30 years (Pagel \& Patchett
\cite{pp75}; Hartwick \cite{har76}; Twarog \cite{twa80}; Edmunds
\cite{edm90}).  Here we follow the approach and the terminology of
Tinsley (\cite{tin80}) and Matteucci (\cite{mat01}), namely we assume
for simplicity {\it linear} flows (we assume gas flows proportional to
the star formation rate (SFR)).  Therefore, the outflow rate $W (t)$
is given by:

\begin{equation}
W (t) = \lambda (1 - R) \psi (t),
\label{eq:w}
\end{equation}
\noindent
where $\psi (t)$ is the SFR, and the infall rate $A (t)$ is given by:

\begin{equation}
A (t) = \Lambda (1 - R) \psi (t).
\label{eq:a}
\end{equation} 
\noindent
Here $\lambda$ and $\Lambda$ are two proportionality constants.  The
first assumption is justified by the fact that, the larger the SFR is,
the more intense are the energetic events associated with it (in
particular supernova explosions and stellar winds), therefore the
larger is the chance of having a large-scale outflow (see e.g. Silk
\cite{silk03}).  A proportionality between $A (t)$ and $\psi (t)$ can
be instead justified by the fact that the infall of gas provides a
continuous reservoir for the star formation.  A careful analysis of
this assumption is given in Sect. \ref{sec:limits}.

The system of equations we need to solve is therefore:

\begin{equation}
\cases{{d M_{tot} \over d t} = (\Lambda - \lambda) (1 - R) \psi (t) \cr
{d M_g \over d t} = (\Lambda - \lambda - 1) (1 - R) \psi (t) \cr
{d M_Z \over d t} = (1 - R) \psi (t) [\Lambda Z_A + y_Z - (\lambda + 1) Z]}
\label{eq:system}
\end{equation}
\noindent
where $M_Z$ is the mass of metals ($M_Z=Z \cdot M_g$) and $Z_A$ is the
metallicity of the infalling gas.  For the moment, $Z_A$ is assumed to
be a constant, although it could change with time due to the pollution
of the intracluster medium (ICM) by means of the metals ejected
through the galactic winds.  This aspect will be considered in detail
in Sect. 3.2 and Sect. 3.3.  With some algebra, this set of equations
can be worked out, yielding:

\begin{equation}
Z = {{\Lambda Z_A + y_Z} \over \Lambda}\biggl\lbrace 
1 - \bigl[ (\Lambda - \lambda) 
- (\Lambda - \lambda - 1) \mu^{-1} \bigr]^{\Lambda \over {\Lambda - \lambda 
- 1}} \biggr\rbrace.
\label{eq:sol}
\end{equation}
\noindent
To find this solution we have made use of the initial conditions
$Z(0)=0$, $M_g (0) = M_{tot} (0) = M_{g, 0}$.  It is worth pointing
out that the explicit dependence of $Z$ on time is hidden in the time
dependence of $\mu$ (see also Sect. \ref{sec:limits}).

\begin{figure}
\centering
\epsfig{file=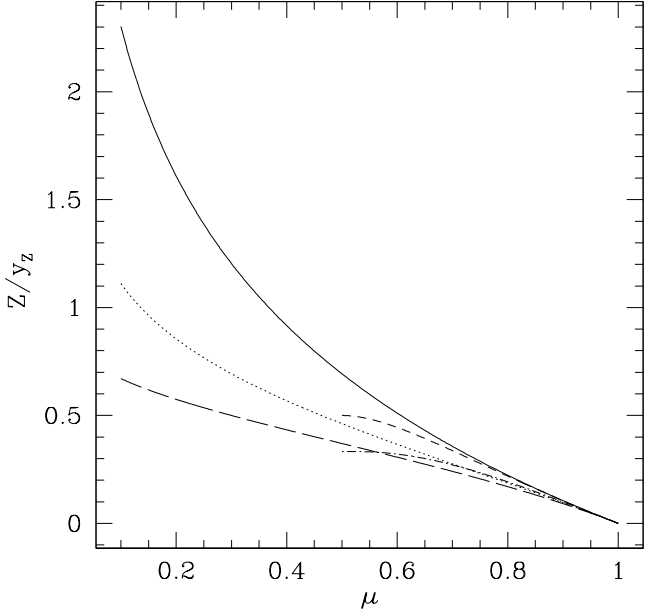, height=9cm,width=9cm}
\caption[]{\label{closedbox} Evolution of the metallicity as a function 
of $\mu$ for the closed box model (solid line) and models with gas flows: 
$\lambda=2$ and $\Lambda=0$ (dotted line); $\lambda=0$ and $\Lambda=2$ 
(short-dashed line); $\lambda=3$ and $\Lambda=1$ (long-dashed line); 
$\lambda=1$ and $\Lambda=3$ (dot-dashed line).  The metallicity of the 
infalling gas $Z_A$ is assumed to be 0.}
\end{figure}

\begin{figure}
\centering
\epsfig{file=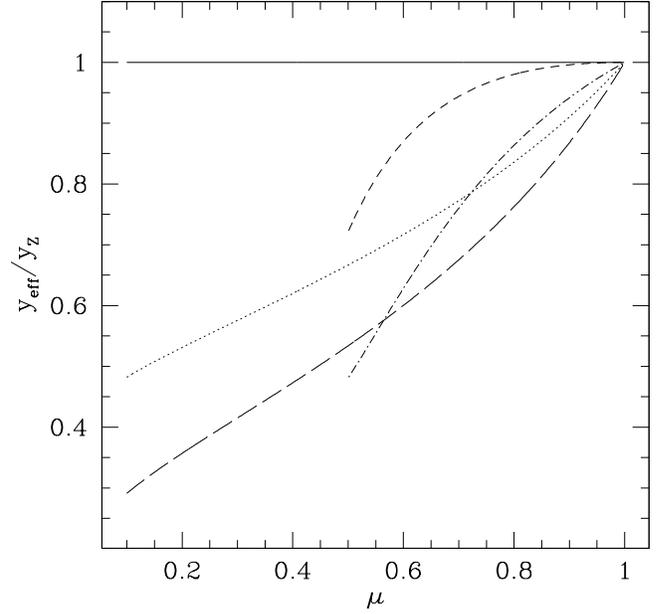, height=9cm,width=9cm}
\caption[]{\label{yeff} Normalized effective yield as a function of $\mu$.  
Notation as in Fig.~\ref{closedbox}}
\end{figure}

We show in Fig.~\ref{closedbox} the evolution of $Z/y_Z$ versus $\mu$
in the case $Z_A=0$ for different values of $\lambda$ and $\Lambda$.
It is worth reminding that $\mu = 1$ at the beginning of the evolution
of the galaxy and its value decreases with time.  Therefore, the time
axis is inverted compared to the $\mu$-axis in this figure and in
similar ones.  We can see that all the models with gas flows produce
metallicities smaller than the ones predicted by the closed box model
(solid line).  This is in agreement with the first and the third of
Edmund's theorems (Edmunds \cite{edm90}), since we have assumed that
the infall gas is pristine ($Z_A=0$).  Fig.~\ref{yeff} shows the
effective yields $y_{eff} = Z / \ln (\mu^{-1})$ (normalized to $y_Z$)
of the same models presented in Fig~\ref{closedbox}.  $y_{eff}$ is
often used in literature to interpret data and to analyze to what
extent the chemical evolution of a galaxy deviates from the behavior
predicted by the closed box model (Garnett \cite{garn02}; Tremonti et
al. \cite{trem04}).  We can notice from eq.  (\ref{eq:sol}) that the
solution diverges for models in which $\lambda > \Lambda$ when $\mu$
approaches 0, therefore, we show $Z/y_Z$ and $y_{eff} / y_Z$ versus
$\mu$ only until $\mu \sim 0.1$.  Indeed, Prantzos \& Aubert
(\cite{pa95}) pointed out that IRA cannot be applied in situations in
which low $\mu$ are obtained, because the amount of gas ejected by low
mass stars at late times can substantially modify the final
abundances, also for elements produced on short timescales by massive
stars.  We notice here also that, in models in which the infall rate
is larger than the outflow rate, not all the values of $\mu$ are
allowed.  Solving the first two equations of the system
(\ref{eq:system}), we get:

\begin{equation}
\mu = {{\Lambda - \lambda - 1} \over {\Lambda - \lambda}} + 
{1 \over {\Lambda - \lambda}} {M_{g, 0} \over M_g}.
\label{eq:mu}
\end{equation}
\noindent
In the models in which $\Lambda > \lambda + 1$, $M_g$ is always
increasing (eq. \ref{eq:system}), therefore $\mu$ ranges between 1 and
a minimum value:

\begin{equation}
\mu_{\rm min} = {{\Lambda - \lambda - 1} \over {\Lambda - \lambda}}.
\label{eq:mumin}
\end{equation}
\noindent
For models in which $\lambda + 1 > \Lambda > \lambda$ there is no
$\mu_{\rm min}$ but there is a upper limit reachable by the gas mass
which is given by $M_{g, lim} = M_{g, 0} / (\lambda + 1 - \Lambda)$.
For this reason we plotted in Figs.~\ref{closedbox} and \ref{yeff} the
model with $\lambda=0$ and $\Lambda=2$ (short-dashed line) and the
model with $\lambda=1$ and $\Lambda=3$ (dot-dashed line) only for $\mu
\geq \mu_{\rm min} = 0.5$.

\section{Simple models with differential winds}
\label{diffwinds}

\subsection{The basic solution}
\label{diffwinds_basic} 

We present in this section the chemical evolution of galaxies in which
a differential wind is assumed, namely in which the metals are more
easily channelled out of the parent galaxy than the pristine gas.  The
easiest way to consider a differential wind in the framework of simple
models of chemical evolution is to assume that the metallicity of the
gas carried out in the galactic wind is proportional to the
metallicity of the ISM with a proportionality constant larger than
one.  If we define $Z^o$ as the metallicity of the outflowing gas,
this condition implies that $Z^o = \alpha Z$ with the ejection
efficiency $\alpha > 1$.  In the metallicity budget (third equation in
(\ref{eq:system})) we have to assume that the negative term due to the
galactic wind is given by $W (t) Z^o = \alpha \lambda (1 - R) \psi
(t)$.  A similar approach has been used by Salvadori et
al. (\cite{sfs08}) but in the framework of a semi-analytical model for
the evolution of galaxies.  With our simple approach we are able to
determine analytical expressions for the evolution of $Z$, which allow
us to understand more clearly the effect of galactic winds on the
chemical evolution of galaxies.

The set of the equations we have to solve in this case is very similar
to (\ref{eq:system}), with the only difference given by the
metallicity budget equation, which we modify as follows:

\begin{equation}
{d M_Z \over d t} = (1 - R) \psi (t) [\Lambda Z_A + y_Z - (\lambda \alpha 
+ 1) Z].
\label{eq:diffw}
\end{equation}
\noindent
The solution of this new set of equations is given by:

\begin{equation}
Z = {{\Lambda Z_A + y_Z} \over {\Lambda + (\alpha - 1)} \lambda}\biggl\lbrace 
1 - \bigl[ (\Lambda - \lambda) 
- (\Lambda - \lambda - 1) \mu^{-1} \bigr]^
{{\Lambda + (\alpha - 1) \lambda} \over {\Lambda - \lambda - 1}} 
\biggr\rbrace.
\label{eq:diffwsol}
\end{equation}
\noindent
It is straightforward to see that we can obtain eq. (\ref{eq:sol}) in
the case $\alpha = 1$ (i.e. in the case in which the galactic wind is
not differential).  We remind here that this solution is valid only
for primary elements.  The evolution of secondary elements (namely
elements synthesized from the metals present in the stars at birth) in
the framework of simple models with gas flows has been extensively
studied by K\"oppen \& Edmunds (\cite{ke99}) and we refer the reader
to this paper to find out more about it.

\begin{figure}
\centering
\epsfig{file=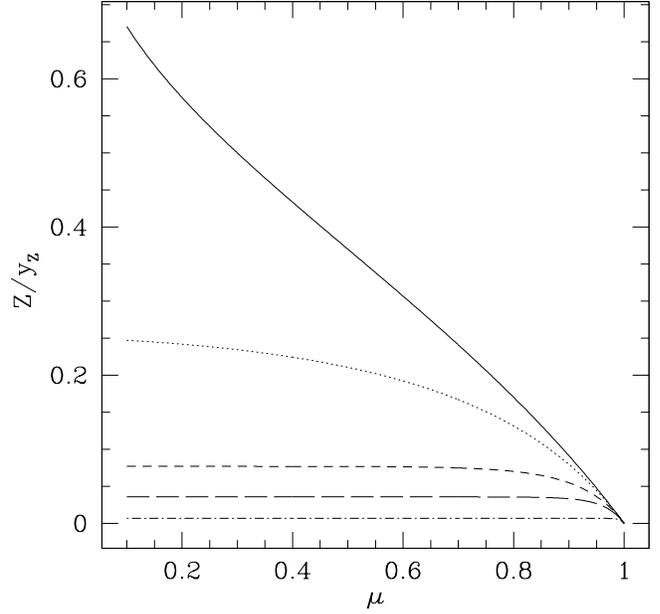, height=9cm,width=9cm}
\caption[]{\label{diff31} Evolution of the metallicity as a function 
of $\mu$ for simple models with differential winds, $\lambda = 3$,
$\Lambda = 1$ and $Z_A = 0$.  Plotted are models with the differential
wind parameter $\alpha=2$ (dotted line), 5 (short-dashed line), 10
(long-dashed line) and 50 (dot-dashed line).  For reference, also the
model with $\alpha = 1$ (normal wind) is plotted (solid line).}
\end{figure}

\begin{figure}
\centering
\epsfig{file=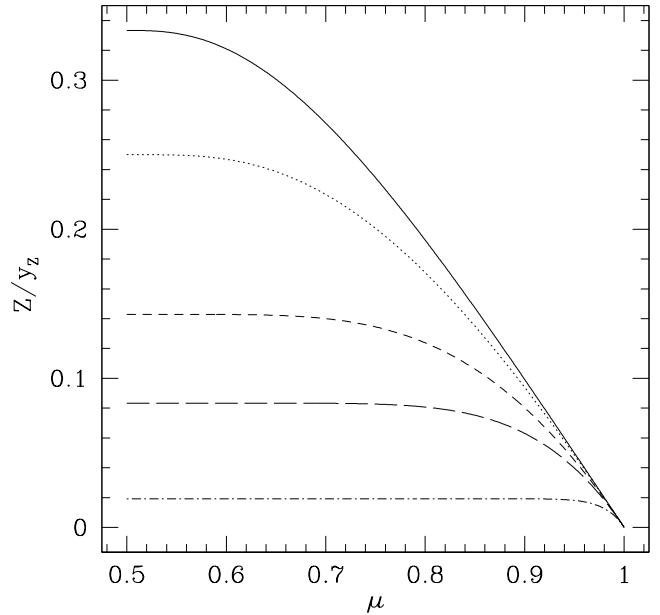, height=9cm,width=9cm}
\caption[]{\label{diff13} As in Fig.~\ref{diff31} but for 
$\lambda = 1$ and $\Lambda = 3$}.
\end{figure}

Assuming again for simplicity an infall of pristine gas (i.e. $Z_A =
0$) we can see how the chemical evolution of the galaxy varies with
the ejection efficiency $\alpha$.  In Figs.~\ref{diff31} and
\ref{diff13} we show Z/$y_Z$ vs. $\mu$ for different values of
$\alpha$ and gas flow parameters ($\lambda$, $\Lambda$) taken as (3,
1) and (1, 3), respectively.  The first couple of values can be
attributed to objects in which galactic winds have a prominent role.
This is usually the case in dwarf galaxies, since the potential well
of these objects is very shallow and large-scale outflows must be
common (Dekel \& Silk \cite{ds86}, but see also Skillman
\cite{ski97}).  The second set of parameters ($\lambda = 1$, $\Lambda
= 3$) can more properly refer to spiral galaxies, although large
outflow episodes in massive spirals are not very likely.  In general,
the effect of differential winds can be more clearly distinguished in
models in which the outflow is prominent compared to the infall.  For
this reason, we will pay from now on more attention on the models for
which $\lambda > \Lambda$, taking $\lambda = 3$, $\Lambda = 1$ as
reference values.  Our aim in this paper is not to describe a
prototypical dwarf galaxy, therefore the combination of parameters
($\lambda$, $\Lambda$) = (3, 1) is purely indicative and serves as a
benchmark of our models with differential winds.

As expected, the differential wind determines a significant reduction
of the metallicity of the galaxy.  It can be shown analytically that,
for each value of $\alpha > 1$, eq. (\ref{eq:diffwsol}) produces
metallicities systematically smaller that the ones obtained from
eq. (\ref{eq:sol}).  Moreover, it is evident from these plots the
asymptotic trend of $Z/y_Z$ as $\mu$ approaches 0 or $\mu_{min}$.
From eq. (\ref{eq:diffwsol}) it is easy to see that, in both cases,
$Z/y_Z$ asymptotically tends to the value $[(\alpha - 1) \lambda +
\Lambda]^{-1}$ provided that $Z_A = 0$, therefore $Z/y_Z$ decreases 
almost linearly with $\alpha$ when $\mu$ is small and $\alpha \gg 1$.

\subsection{Instantaneous ICM mixing models}

In our calculations we have so far assumed that the metallicity of the
infalling gas $Z_A$ is constant in time.  This assumption is not
necessarily true, in particular when differential winds are taken into
account.  In fact, the galactic winds can pollute the surrounding
medium and the infalling gas can have a metallicity increasing with
time.  The increase of the iron abundance between redshift $z \sim 1$
and $z \sim 0$ in the ICM has been recently suggested (Balestra et
al. \cite{bal07}) and can be accounted for in numerical models in
which the chemical evolution of the ICM is affected by galactic winds
and stripping (Calura et al. \cite{cmt07}).  A simple hypothesis we
can make about the variation of the ICM metallicity with time is that
the metals, expelled out of the galaxy through galactic winds,
instantaneously and uniformly mix with the surrounding ICM.  This
approach has been used by detailed chemical evolution models of the
ICM (e.g. Matteucci \& Vettolani \cite{mv88}; Moretti et
al. \cite{mpc03}).  Of course the instantaneous mixing of the
outflowing gas with the whole ICM is rather unlikely.  Nevertheless,
this assumption is useful for illustrative purposes.  This hypothesis
leads to a new set of equations in which, besides $Z_A$, we have to
consider also the variation, although small, of the ICM mass $M_{ICM}$
due to galactic winds and infall.  These 4 equations are:

\begin{equation}
\cases{{d M_g \over d t} = (\Lambda - \lambda - 1) (1 - R) \psi (t) \cr
{d M_Z \over d t} = (1 - R) \psi (t) [\Lambda Z_A + y_Z - 
(\lambda \alpha + 1) Z]\cr
{d M_{Z_A} \over d t} = (1 - R) \psi (t) [\lambda \alpha Z - \Lambda Z_A]\cr
{d M_{ICM} \over d t} = (1 - R) \psi (t) (\lambda - \Lambda),
}
\label{eq:newsystem}
\end{equation}
\noindent
where $M_{Z_A}$ is the mass of metals in the ICM ($M_{Z_A}=Z_A \cdot
M_{ICM}$).  In order to solve this set of equations we have to define
the ratio between gas mass and ICM mass $\eta = M_g / M_{ICM}$ and a
new parameter $k = M_{g, 0} / M_{ICM} (0)$, where $M_{ICM} (0)$ is the
initial ICM mass.  Once we have defined the values $X_a$ and $X_b$
such that:

\begin{equation}
X_a^{-1} = {{\lambda - \Lambda} \over {(k + 1) (\lambda - \Lambda) + 1}}
\eta - {{\Lambda - \lambda -1} \over {(k + 1) (\lambda - \Lambda) + 1}},
\end{equation}
\noindent
and

\begin{equation}
X_b^{-1} = {{k (\lambda - \Lambda)} \over {(k + 1) (\lambda - \Lambda) + 1}}
- {{k (\Lambda - \lambda -1)} \over {(k + 1) (\lambda - \Lambda) + 1}} 
\eta^{-1},
\end{equation}
\noindent
the solution of the system of equations (\ref{eq:newsystem}) is given
by:

\begin{equation}
Z = {y_Z \over {\Lambda \alpha X_a^{\lambda \over {\Lambda - \lambda}}
- (\Lambda - \lambda) (\alpha - 1)}} \biggl[
1 - X_b^{-{{\Lambda \alpha X_a^{\lambda \over {\Lambda - \lambda}}
- (\Lambda - \lambda) (\alpha - 1)} \over {\Lambda - \lambda -1}}}
\biggr].
\label{eq:complsol}
\end{equation}

\begin{figure}
\centering
\epsfig{file=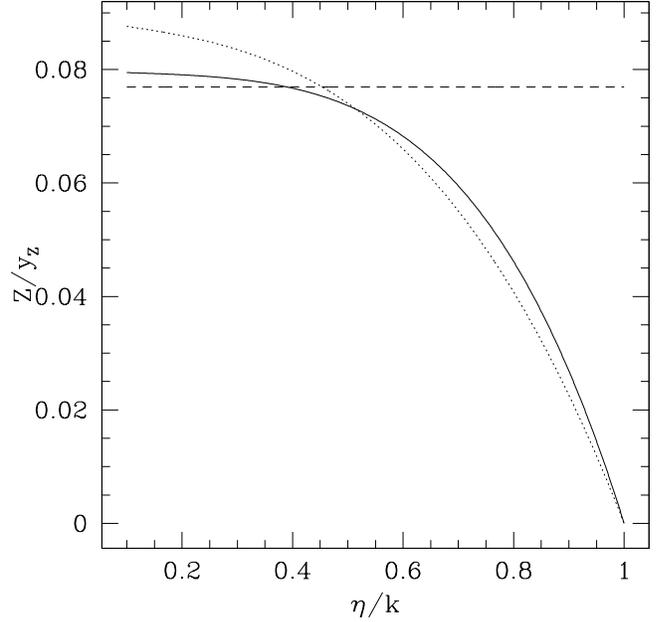, height=9cm,width=9cm}
\caption[]{\label{fig:compl} Chemical evolution of models in which 
the metallicity of the ICM varies with time (instantaneous ICM mixing
models) and with initial mass ratio $k$ (see text) equal to 0.1 (solid
line) and 0.5 (dotted line).  Also shown in this plot is the
asymptotic trend of the simple differential wind model for $\mu$
approaching 0 (dashed line).  For all models, $\lambda = 3$, $\Lambda
= 1$ and $\alpha = 5$.}
\end{figure}

It is straightforward to see that eqs (\ref{eq:complsol}) and
(\ref{eq:diffwsol}) produce the same results if the infall is
negligible (i.e. if $\Lambda \sim 0$).  Considering again a model with
$\Lambda = 1$ and $\lambda = 3$ we can show the trend of $Z/y_Z$ in
the case $\alpha = 5$ for different values of $k$.  We notice first
that, unlike $\mu$, $\eta$ does not monotonically decrease, but its
variation depends on the values of $\lambda$ and $\Lambda$.  In the
present example, since $\lambda > \Lambda$, $M_g$ decreases with time
and $M_{ICM}$ increases with time, therefore $\eta$ ranges between $k$
at the beginning and 0 during the late stages of the evolution of the
galaxy.  We show in Fig.~\ref{fig:compl} the evolution of $Z/y_Z$ as a
function of $\eta/k$ for $k=0.1$ (solid line) and $k=0.5$ (dotted
line).  The model with $k=0.5$ predicts larger abundances because the
metals carried out by the galactic wind are diluted with less ICM,
therefore $Z_A$ is larger.  These results cannot be compared to the
results of eq. (\ref{eq:diffwsol}) in a straightforward manner since
they are expressed in different units.  However, we show in this plot
the asymptotic trend of the simple differential wind model
(short-dashed line in Fig.~\ref{diff31}) for $\mu$ approaching 0
(dashed line).  This comparison allows us to notice that, as expected,
eq. (\ref{eq:complsol}) predicts larger metallicities compared to
models with $Z_A = 0$.  However, as long as $k < 1$, the differences
are small (of the order of $\sim$ 0.1 - 0.2 dex).

We can also calculate the ICM metallicity, which turns out to be:

\begin{equation}
Z_A = \alpha Z \bigl(1 - X_a^{\lambda \over {\Lambda - \lambda}} 
\bigr).
\label{eq:icmz}
\end{equation}
\noindent
In the case we are considering more closely, $\lambda = 3$ and
$\Lambda = 1$, therefore the ICM mass increases with time and $X_a$
ranges between 1 and $(1 + k)$.  The larger value $Z_A/Z$ can attain
is therefore:

\begin{equation}
{Z_A \over Z} = \alpha \biggl[ 1 - \bigl( 1 + k \bigr)^{\lambda \over 
{\Lambda - \lambda}} \biggr],
\label{eq:zratio}
\end{equation}
\noindent
which is smaller than 1 provided that:

\begin{equation}
k < k_{lim} = \biggl[{\alpha \over {\alpha - 1}} \biggr]^{{\lambda - 
\Lambda} \over \lambda} - 1.
\label{eq:klim}
\end{equation}
\noindent
Once the ratio $Z_A/Z$ between ICM and ISM metallicity is
observationally known, eq. (\ref{eq:zratio}) can be used to constrain
$\alpha$.

\subsection{Galactic fountain models}

A special case of variable infall metallicity is represented by the
situation in which $Z_A = Z^o = \alpha Z$, namely the metallicity of
the infalling gas is set to be always equal to the one of the galactic
wind.  This condition implies therefore that the very same gas that
has been driven out of the galaxy by energetic events can subsequently
rain back to the galaxy, due to the pull of its gravitational
potential.  This kind of duty cycle is well known in astrophysics and
it has been named {\it galactic fountain} (Shapiro \& Field
\cite{sf76}; Bregman \cite{breg80}).  In order to solve the chemical
evolution in this special case, we have necessarily to assume that
$\lambda \geq \Lambda$ since the reservoir for the infall gas is given
by the gas expelled out of the galaxy through galactic winds.  The
solution of the chemical evolution in the presence of galactic
fountains is given by:

\begin{equation}
Z = {y_Z \over {(\lambda - \Lambda) (\alpha - 1)}}\biggl\lbrace 
1 - \bigl[ (\Lambda - \lambda) 
- (\Lambda - \lambda - 1) \mu^{-1} \bigr]^
{{(\lambda - \Lambda) (\alpha - 1)} \over {\Lambda - \lambda - 1}} 
\biggr\rbrace.
\label{eq:fountsol}
\end{equation}

\begin{figure}
\centering
\epsfig{file=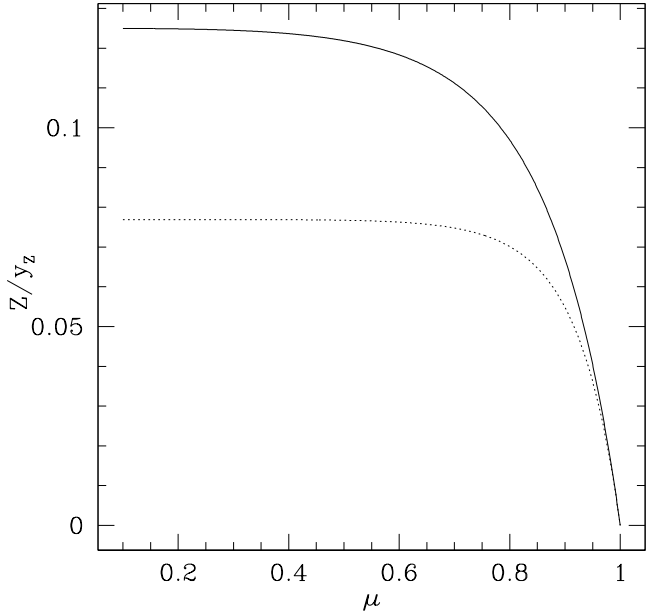, height=9cm,width=9cm}
\caption[]{\label{fount} Comparison of the chemical evolution of a 
galactic fountain model (solid line) with a model with infall of
pristine gas ($Z_A=0$, dotted line).  For both models, $\lambda = 3$,
$\Lambda = 1$ and $\alpha = 5$.}
\end{figure}

A comparison of this solution with the solution of eq.
(\ref{eq:diffwsol}) with infall of pristine gas ($Z_A = 0$) is shown
in Fig.~\ref{fount}.  For both models plotted in this figure we have
assumed $\lambda = 3$, $\Lambda = 1$ and $\alpha=5$.  As expected, the
galactic fountain model predicts larger vales of $Z$.  These values
exceed also the metallicities attained by the models with
instantaneous mixing of the ICM (Sect. 3.3) shown in
Fig.~\ref{fig:compl}.  This result is expected because the galactic
fountain model maximizes the enrichment in metallicity of the
infalling gas, whereas in the models shown in Sect. 3.3 the metals
escaping from the galaxy are diluted in a large amount of ICM.  From
eqs. (\ref{eq:diffwsol}) and (\ref{eq:fountsol}) we can also notice
that the metallicity ratio tends asymptotically to the value $[(\alpha
- 1) \lambda + \Lambda] / [(\lambda - \Lambda) (\alpha - 1)]$ which is
always larger than one (for the special choice of parameters it is
equal to 13/8) but it tends to 1 for $\alpha \gg 1$.  In our example,
only an infall of pre-enriched gas, whose metallicity is larger than
5/8 $y_Z$ can therefore produce a metallicity in the galaxy larger
than the one attained by the galactic fountain model.  From
eq. (\ref{eq:fountsol}) we can also see that, in the special case in
which $\lambda = \Lambda$, $Z/y_Z$ tends to the solution of the closed
box model (eq. \ref{eq:simple}).  This is due to the fact that, in the
framework of the simple models of chemical evolution, having outflow
and infall with the same rate and metallicity or not having gas flows
at all is formally the same.  Indeed, the fountains take a finite and
non-negligible time to orbit around and fall back to the galaxy
(Spitoni et al. \cite{srm08}).  This implies a delay in the mixing of
metals in the ISM, which conflicts with the fourth assumption of the
simple models of chemical evolution (Sect. 2), therefore only detailed
numerical models can ascertain the effect of this delay on the
chemical evolution of galaxies (Spitoni et al. 2008, in preparation).

\subsection{Limits of the approximations of linear flows}
\label{sec:limits}

In the previous sections we have assumed that both outflow and infall
rates are proportional to the SFR and we have seen that in this way we
obtain solutions in which $\psi (t)$ cancels out.  In this section we
analyze how reliable are these assumptions and what kind of results we
obtain if we assume general infall and outflow laws (see also Edmunds
\cite{edm90}; K\"oppen \& Edmunds \cite{ke99}).  First of all we
notice that a correlation between infall rate and SFR arises naturally
because the larger is the amount of infalling gas, the higher is the
reservoir of gas inside the galaxy available to form stars.  An
example is given by detailed models of the chemical evolution of the
Milky Way (Chiappini et al. \cite{cmg97}; Cescutti et
al. \cite{cescu07}; Colavitti et al. \cite{cmm08}) in which two main
phases of gas infall turn out to produce two distinct episodes of star
formation (although the similar behavior between the two rates does
not imply that their ratio is constant).  We will show in this section
that, under reasonable assumptions, a constant ratio between SFR and
infall rate can be attained during the late evolution of the
considered galaxies.  On the other hand, a proportionality between the
outflow rate and the SFR is quite realistic and it has been shown both
observationally (e.g. Heckman \cite{heck02}) and theoretically
(e.g. Silk \cite{silk03}).

Therefore, let us assume for simplicity that eq. (\ref{eq:w}) still
holds and that the infalling gas is pristine (e.g. $Z_A = 0$).  The
extension of our calculations to the case in which $W (t)$ is a
generic function is straightforward.  If we assume a generic infall
law $A (t)$, the system of equations we need to solve is:

\begin{equation}
\cases{{d M_{tot} \over d t} = A (t) - \lambda (1 - R) \psi (t) \cr
{d M_g \over d t} = A (t) - (\lambda + 1) (1 - R) \psi (t) \cr
{d M_Z \over d t} = (1 - R) \psi (t) [y_Z - (\alpha \lambda + 1) Z (t)],}
\label{eq:system_general}
\end{equation}
\noindent
subject to the usual initial conditions $Z(0)=0$, $M_g (0) = M_{tot}
(0) = M_{g, 0}$.  In this case, however, it is not possible to solve
this system of equations canceling out $\psi (t)$.  Therefore, at
variance with what we have done so far, we must assume some dependence
of $\psi (t)$ on $M_g$ (Schmidt law).  We assume a linear Schmidt law
(e.g. $\psi (t) = S M_g (t)$), because it is the only formulation for
which the results can be expressed analytically.  For Schmidt laws
with different exponents detailed numerical models are required.  The
equation for the evolution of the gas mass can be simply solved,
yielding:

\begin{equation}
M_g (t) = e^{- (\lambda + 1) (1 - R) S t} \biggl[ \int_0^t A (u) 
e^{(\lambda + 1) (1 - R) S u} du + M_{g, 0} \biggr].
\label{eq:mg_generic}
\end{equation}
\noindent
Substituting this expression into the third equation of the system
(\ref{eq:system_general}) also the evolution of the metallicity as a
function of time can be calculated.  The result is the following:

\begin{equation}
Z (t) = (1 - R) S y_Z \bigl[I (t) \bigr]^{-1} \int_0^t I (u) du,
\label{eq:z_gen}
\end{equation}
\noindent
where

\begin{equation}
I (t) = e^{\lambda (1 - R) S (\alpha - 1) t + \int_o^t {{A (u) 
e^{(\lambda + 1) (1 - R) S u}} \over {M_{g, 0} + \int_0^u A (v)
e^{(\lambda + 1) (1 - R) S v} dv}} du}.
\end{equation}

We consider now as an example an exponential infall law, study what
kind of metallicity evolution it produces and compare the results
with those presented in Sect. \ref{diffwinds_basic}, bearing in mind
that, in the present case, we have to define the value of some
constants ($M_{g, 0}$, $S$ as well as the constants defining the
infall rate), whereas in the previous case it was enough to define the
proportionality constant $\Lambda$.  We have:

\begin{equation}
A (t) = K e^{- t / t_i} = K e^{- \tau},
\end{equation}
\noindent
where $t_i$ is a characteristic infall timescale and $\tau = t / t_i$.
We can calculate $K$ assuming that the initial gas mass $M_{g, 0}$ is
proportional to the total amount of infalling gas through a
proportionality constant $\gamma$ (namely we assume that $M_{g, 0} = \gamma
\int_0^\infty A (t) dt$).  In this way, we obtain $K = {M_{g, 0} \over
{\gamma t_i}}$.  Eq. (\ref{eq:mg_generic}) can be therefore solved:

\begin{equation}
M_g (\tau) = {M_{g, 0} \over K_1} \Biggl[ e^{- \tau} + K_1 
e^{- (\lambda + 1) (1 - R) S t_i \tau}
\Biggr],
\label{eq:mg_exp}
\end{equation}
\noindent
where $K_1$ is a constant (see the appendix for its exact value).

\begin{figure}
\centering
\epsfig{file=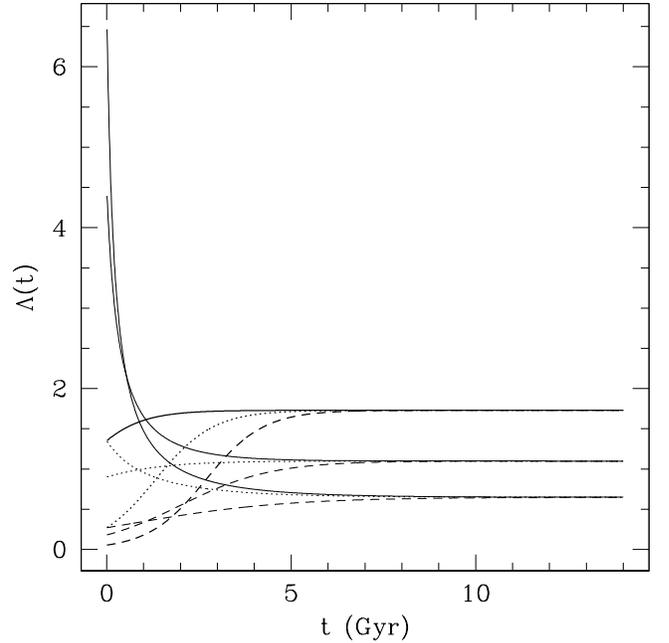, height=9cm,width=9cm}
\caption[]{\label{infall_sfr} Ratio between the infall rate and the 
SFR as a function of time for models with an exponential infall and
different $t_i$ and $\gamma$ (see text).  Heavy lines: $t_i = 5$ Gyr;
normal lines: $t_i = 1.5$ Gyr; light lines: $t_i = 1$ Gyr.  Solid
lines: $\gamma = 0.2$; dotted lines: $\gamma = 1$; dashed lines:
$\gamma = 5$.}
\end{figure}

We can now evaluate the ratio $\Lambda (t) = {A (t) \over {(1 - R)
    \psi (t)}}$ in order to test if and under which conditions we can
still assume a proportionality between the SFR and the infall rate $A
(t)$.  In Fig. \ref{infall_sfr} we show the resulting $\Lambda (t)$
for different values of $\gamma$ (0.2, 1 and 5) and $t_i$ (1, 1.5 and
5 Gyr).  In order to draw these curves we have assumed $\lambda = 3$
(in agreement with what assumed in Sect. \ref{diffwinds_basic}) and $S
= 1$ Gyr$^{-1}$.  This is an upper limit for the star formation
efficiency in dwarf galaxies (see. e.g. Lanfranchi \& Matteucci
\cite{lm03}).  $R$ is assumed to be 0.26, in agreements with
the nucleosynthetic calculations of Woosley \& Weaver (\cite{ww95})
assuming a Salpeter IMF.  A small value of $\gamma$ implies
that the galaxy initially does not have a large amount of gas and the
star formation is mostly triggered by the infalling gas, therefore the
infall rate initially prevails over the SFR.  A large $t_i$ implies a
large infall rate at late times, therefore $\Lambda (t)$ tends
asymptotically to a large value.  In general the behavior of $\Lambda
(t)$ depends on a complex interplay between $\gamma$, $t_i$, $S$ and
$\lambda$ (see also K\"oppen \& Edmunds \cite{ke99}) but we can see
that under reasonable assumptions the models show a first phase
(lasting a few Gyr) in which $\Lambda (t)$ varies with time, but then
it stabilizes to a constant value.  As a general result, only if
we assume infall timescales much shorter than the star formation
timescale (i.e. the inverse of $S$) $\Lambda (t)$ shows significant
variations over the whole evolution of the galaxy, although these
variations are milder if a large value of $\lambda$ is assumed.  For
instance, if we take $S = 0.1$ Gyr$^{-1}$, a constant $\Lambda (t)$ is
attained by assuming either large values of $\lambda$ or large values
of $t_i$ or values of $\gamma$ such that $K_1 \sim 0$.  If we assume a
quadratic Schmidt law, the infall timescales required to obtain a
constant $\Lambda (t)$ tend to be larger than in the case of a linear
Schmidt law.  We can conclude that a linear flow can be a reasonable
approximation of the late evolution of a galaxy provided that the
infall timescale is of the same order of magnitude of the star
formation timescale.

Now we can answer the following question: does an early phase of
variable $\Lambda (t)$ significantly affect the evolution of the
metallicity?  To do that we have to solve eq. (\ref{eq:z_gen}).  The
resulting evolution of $Z / y_Z$ is given by:

\begin{equation}
{Z (\tau) \over y_Z} = {{K_2 + K_3 e^{E_1 \tau} - \bigl(K_2 + K_3\bigr) 
e^{- E_2 \tau}} \over {e^{E_1 \tau} + K_1}},
\end{equation}
\noindent
(see the Appendix for the expressions of the involved constants).

\begin{figure}
\centering
\epsfig{file=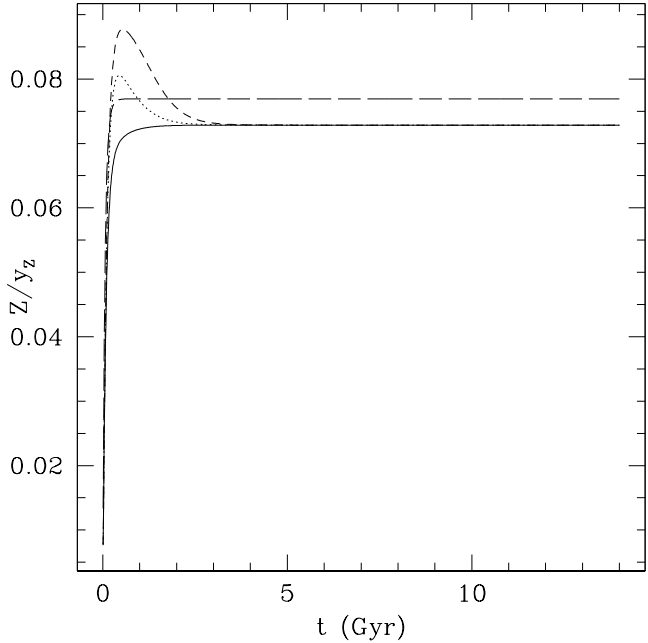, height=9cm,width=9cm}
\caption[]{\label{z_num} Evolution of the metallicity as a function 
of time for models with an exponential infall and different $\gamma$.
Solid line: $\gamma = 0.2$; dotted line: $\gamma = 1$; short-dashed
line: $\gamma = 5$.  For all the models it is assumed $t_i = 1.5$ Gyr.
Also plotted (long-dashed line) the evolution of a model in which a
linear infall (with $\Lambda = 1$) is assumed.  For all models $\alpha
= 5$.}
\end{figure}

We show in Fig. \ref{z_num} the evolution of $Z (t) / y_Z$ as a
function of time for different values of $\gamma$.  In these models we
assume $t_i = 1.5$ Gyr because this value leads to a $\Lambda$ close
to 1 (see Fig. \ref{infall_sfr}).  To compare these models with the
results shown in Sect. \ref{diffwinds_basic} we combine the first two
equations of the system (\ref{eq:system}) to get $\mu (t)$ in a model
with linear flows and a linear Schmidt law.  By means of
eq. (\ref{eq:diffwsol}) we then obtain the evolution of $Z$ as a
function of time, which turns out to be:

\begin{equation}
{Z (t) \over y_Z} = 
{{1 - \bigl[\bigl(\Lambda - \lambda + 1\bigr) - 
\bigl( \Lambda - \lambda\bigr) e^{(\lambda + 1 - \Lambda) (1 - R) S t}
\bigr]^{{\Lambda + (\alpha - 1) \lambda} \over {\Lambda - \lambda - 1}}}
\over 
{\Lambda + (\alpha - 1) \lambda}}.
\end{equation}
\noindent
We plotted also this curve in Fig. \ref{z_num} for $\Lambda = 1$
(long-dashed line).  The difference with the models with exponential
infall laws is very small and mostly due to the fact that the
asymptotic value of $\Lambda (t)$ for the exponential infall models is
not exactly 1.  Therefore we can conclude that the results we have
shown in Sect. \ref{diffwinds_basic} do not change substantially if we
assume an exponential infall instead of an infall rate proportional to
the SFR.  This demonstrates also that the major source of error in the
solutions of the simple models is the IRA assumption rather than the
assumption of linear flows.

\section{The abundance ratios}

From eq. (\ref{eq:diffwsol}) it is straightforward to determine the
abundance ratio between two chemical elements $i$ and $j$ assuming
that the yields do not depend on metallicity (see also
Sect. \ref{simple}), namely:

\begin{equation}
{Z_i \over Z_j} = {{y_i + \Lambda Z_{A, i}} \over {y_j + 
\Lambda Z_{A, j}}}.
\label{eq:abratio}
\end{equation}
\noindent
This abundance ratio is therefore determined uniquely by constants
which depend on the IMF and on the properties of the infalling gas.
In the models with variable infall metallicities, $Z_i/Z_j$ is simply
given by the yields ratio $y_i/y_j$ (see eqs. \ref{eq:complsol} and
\ref{eq:fountsol}).  With our simplifying assumptions, the abundance
ratio must remain constant throughout the whole evolution of the
galaxy.  Also the abundance ratios of the ICM (see eq. \ref{eq:icmz})
must remain constant with $\eta$ (and therefore with time).  The
constancy of abundance ratios in the framework of simple models of
chemical evolution was also pointed out by Matteucci \& Chiappini
(\cite{mc05}), in which the authors showed that variations of the
abundance ratios can be accounted for only by relaxing the IRA and by
properly taking into account stellar lifetimes.  Here we outline that
this result holds also if we consider differential winds.  

By neglecting the contribution of SNeIa, it is expected that the
[O/Fe] ratio does not change with time (and therefore with [Fe/H])
since, in this case, both elements are produced by SNeII, on similar
timescales.  Indeed, detailed models of the evolution of the Galaxy
(Chiappini et al. \cite{cmg97}) show a mild decrease of the predicted
(and observed) [O/Fe] as a function of [Fe/H] even without considering
SNeIa, due to the variation of the yields with stellar mass, but this
can account for only $\sim$ 0.2 dex of the [O/Fe] variation.  A much
more significant reduction of the [O/Fe] can be attained by taking
into consideration SNeIa, which provide most of the Fe of the ISM and
explode with a delay compared to the SNeII.  This model (also called
{\it time-delay model}) remains the most plausible explanation of the
variation of [O/Fe] vs. [Fe/H] in various galaxies (Matteucci \&
Greggio \cite{mg86}; Lanfranchi \& Matteucci \cite{lm03}). In
particular, Lanfranchi \& Matteucci (\cite{lm03}) interpreted the
steep decline of the [$\alpha$/Fe] ratios at low [Fe/H] (at [Fe/H]
$\sim$ -1.6 dex), observed in dwarf spheroidals of the Local Group
(e.g. Shetrone et al. \cite{scs01}, Venn et al. \cite{venn04}) as due
to the slow SFR acting in these systems, coupled with very efficient
galactic winds.  In this situation, in fact, the Fe production from
Type Ia SNe occurs when the [Fe/H] is still low.  At this point, the
galactic winds extract both $\alpha$-elements and Fe from the galaxy
but while O is less and less produced because of the decrease in the
star formation due to the wind, Fe continues to be produced by Type Ia
SNe.  Unfortunately, SNeIa cannot be treated by means of simple model
(SNeIa arise from CO white dwarfs in binary stars, with non-negligible
lifetimes, therefore the IRA cannot be applied).  Because of that, it
is not possible to solve analytically chemical evolution equations
which include SNeIa and this is the reason why most people make use of
numerical simulations.  However, IRA provides a good approximation of
the evolution of $\alpha$ elements, promptly released by SNeII.
Recently, Savadori et al. (\cite{sfs08}) proposed an alternative
scenario in which the drop in the [O/Fe] is due to strong differential
winds, combined with the relaxation of the IRA, without invoking
SNeIa.  In their model, the star formation is dominated by an early,
very short ($\sim$ 100 Myr) burst.  However, a non-negligible fraction
of SNeIa seems to appear also within 0.1 Gyr in galaxies (see Mannucci
et al. \cite{mann05};
\cite{mann06}), therefore neglecting SNeIa in their model could be
incorrect.  This might be the reason why these authors do not
reproduce accurately the high metallicity tail of the metallicity
distribution function of Sculptor (namely the stars with [Fe/H] $\sim
-1$; see their figs. 4 and 6).  On the other hand, neglecting Type Ia
SNe appears consistent with the stellar lifetimes they have assumed.
These lifetimes predict 150 Myr for a 8 M$_\odot$ star, instead of
30--40 Myr as commonly adopted.  These long lifetimes, together with
the short duration of the main episode of SF and the relaxation of IRA
help in producing the decrease of [O/Fe] since in this case the
evolution of abundance ratios resembles that of an instantaneous burst
in which the [O/Fe] at a given time $t$ follows the ratio of O and Fe
yields of the star dying at that time.

One way to attain a variable [O/Fe] ratio in the framework of simple
models of chemical evolution is by assuming selective winds, namely
that the two elements have different ejection efficiencies.  It has
been suggested since a long time that $\alpha$-elements can be
channelled out of a galaxy more easily than other elements (Marconi et
al. \cite{mmt94}; Martin et al. \cite{mkh02}) but Recchi et
al. (\cite{rmd01}) showed that, during the very first burst of star
formation experienced by a dwarf galaxy, iron can be expelled more
easily than oxygen.  The hypothesis of different ejection efficiencies
$\alpha$ for O and Fe is therefore a reasonable assumption that can be
tested by our models.

\begin{figure}
\centering
\epsfig{file=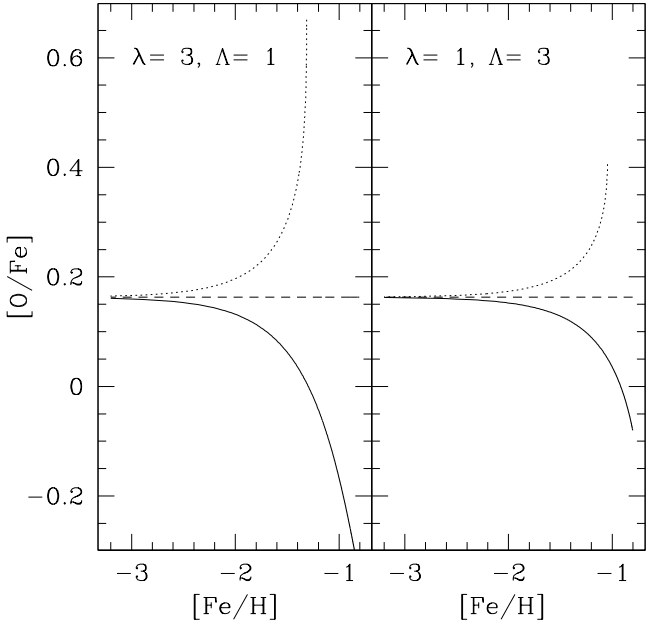, height=9cm,width=9cm}
\caption[]{\label{ofe} [O/Fe] vs. [Fe/H] for models characterized by 
different ejection efficiencies for O and Fe $\alpha_O$ and
$\alpha_{Fe}$ .  ($\alpha_O$, $\alpha_{Fe}$) are (5, 2) (solid line),
(2, 5) (dotted line) and (2, 2) (dashed line).  ($\lambda,
\Lambda$) are (3, 1) (left panel) and (1, 3) (right panel).}
\end{figure}

We assume two different values for these ejection efficiencies
$\alpha_O$ and $\alpha_{Fe}$ and, by means of eq. (\ref{eq:diffwsol}),
we calculate how [O/Fe] varies with [Fe/H].  The results would not
change significantly if we assume eqs. (\ref{eq:complsol}) or
eq. (\ref{eq:fountsol}) instead of eq. (\ref{eq:diffwsol}).  In
Fig.~\ref{ofe} we show for instance the comparisons of models for
which ($\alpha_O$, $\alpha_{Fe}$) = (5, 2) (solid line), (2, 5)
(dotted line) and (2, 2) (dashed line).  We show the results for our
reference set of parameters ($\lambda = 3$ and $\Lambda = 1$) in the
left panel.  To study the dependence of these curves on the inflow and
outflow parameters, we also plot the results for $\lambda = 1$ and
$\Lambda = 3$ (right panel).  In order to calculate the yields, we
have assumed a Salpeter IMF between 0.1 and 40 M$_\odot$ and the
nucleosynthetic prescriptions of Woosley \& Weaver (\cite{ww95}) for
solar metallicity (case B).  For the solar abundances we adopt the
Anders \& Grevesse (\cite{ag89}) values.  We use this set of solar
abundances in order to be consistent with the choice of Woosley \&
Weaver's (\cite{ww95}) yields.

As expected, as the galaxy ages (and, therefore, [Fe/H] increases),
[O/Fe] bends down if the oxygen ejection efficiency is larger (solid
line), whereas it starts increasing if $\alpha_{Fe} > \alpha_O$
(dotted line).  The model with $\lambda = 1$ and $\Lambda = 3$ shows
the same behavior but with reduced variations of the abundance ratios.
This is due to the fact that the difference between the metallicities
attained for $\alpha = 2$ and $\alpha = 5$ in the case ($\lambda$,
$\Lambda$) = (3, 1) is larger than in the case ($\lambda$, $\Lambda$)
= (1, 3) (see Figs. \ref{diff31} and \ref{diff13}).  Of course, the
model in which the ejection efficiencies are the same (dashed line)
does not show any variation of [O/Fe] with [Fe/H] (see
eq. \ref{eq:abratio}).  The [O/Fe] attained by this model ($\sim$ 0.2)
is lower than the [O/Fe] of halo stars and of the most metal-poor
stars of Local Group dwarf galaxies, whose chemical enrichment should
have been dominated by SNeII ejecta.  Indeed, it is known since quite
some time that Woosley \& Weaver's yields overpredict iron (Timmes et
al. \cite{tww95}; Chiappini et al. \cite{cmg97}).  As we have said,
the model with $\lambda = 3$ and $\Lambda = 1$ can be representative
of dwarf spheroidal galaxies.  Indeed, in this model the metallicity
remains significantly lower than the solar value and, if O is ejected
more easily than Fe, a knee at [Fe/H] $\sim$ -1.5 appears, in
agreement with observations (see e.g. Venn et al. \cite{venn04}).  We
point out once more that we are neglecting SNeIa therefore, in the
framework of simple models, O and Fe are both produced by massive
stars, on similar timescales.  As a consequence, they should share a
similar fate and have comparable ejection efficiencies.  Recchi et al.
(\cite{rec04}), by means of detailed chemodynamical simulations, found
similar values of $\alpha_O$ and $\alpha_{Fe}$ when complex star
formation histories are assumed.  Therefore, simple models predict
that variations in the [O/Fe] vs.  [Fe/H] plot cannot be attributed to
differential winds and SNeIa must necessarily be considered in order
to properly interpret this diagram.

\begin{figure}
\centering
\epsfig{file=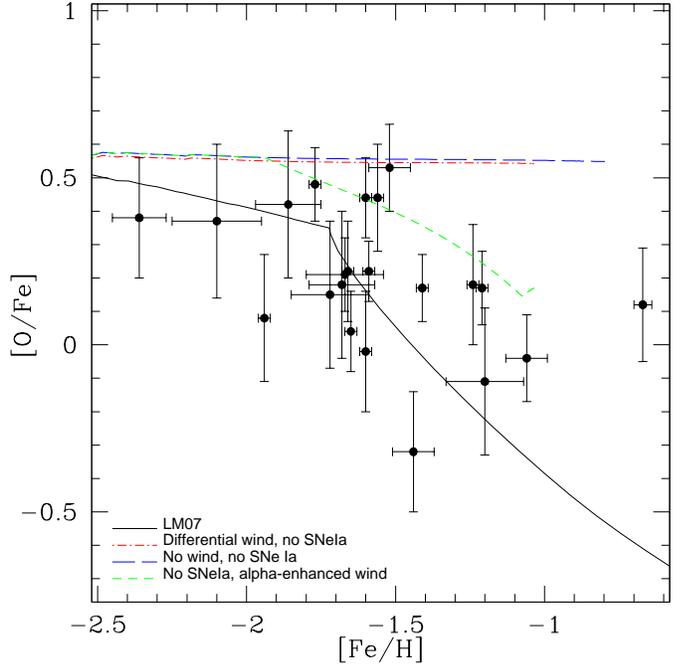, height=9cm,width=9cm}
\caption[]{\label{ofe2} [O/Fe] vs. [Fe/H] for numerical models of a
  standard dwarf spheroidal galaxy.  Solid line: wind
  model like in Lanfranchi \& Matteucci (\cite{lm07});
  dot-short-dashed line: the same model without SNeIa; long-dashed
  line: the same model without SNeIa and without galactic winds; 
  short-dashed line: model without SNeIa in which the
  ejection efficiency of $\alpha$-elements is enhanced.  Data are
  from Bonifacio et al. (\cite{boni00}; \cite{boni04}); Shetrone et
  al.  (\cite{scs01}; \cite{shet03}); Venn et al. \cite{venn04};
  Sadakane et al. (\cite{sada04}); Fulbright et al. (\cite{fulb04});
  Geisler et al. (\cite{geis05}).  For clarity purposes, the 
  dot-short-dashed line has been shifted of 0.01 dex since otherwise 
  it would have been overlapped with the long-dashed line.
}
\end{figure}

Would this conclusion still hold if we relax IRA and take properly
into account stellar lifetimes?  In order to test that, we have
performed a series of numerical simulations of the chemical evolution
of a dwarf spheroidal galaxy whose structural parameters resemble
those of Local Group such as Carina, Draco, Ursa Minor, Sculptor.  In
particular we have run a simulation very similar to the ``standard''
model of Lanfranchi \& Matteucci (\cite{lm03}).  The standard model is
adjusted to reproduce the average [$\alpha$/Fe] and neutron capture
abundance ratios of six local dwarf spheroidals.  It is assumed that
the galaxy forms through the fast and continuous infall of pristine
gas until a mass of $\sim 10^8$ M$_{\odot}$ is reached. The star
formation consists of one long episode (8 Gyr) with low efficiency
($\nu =$ 0.05 Gyr$^{-1}$) and is affected by very intense galactic
winds (with a rate 2 times higher than the SFR).  As the galactic wind
develops, it removes a large gas fraction thus decreasing the SFR.
This effect, coupled with the low star formation rate and the
injection of Fe into the ISM by SNe Ia, gives rise to the knee
observed in [$\alpha$/Fe] and [r-process/Fe].  The models we show now
contain non-selective winds.  We plotted the [O/Fe] vs. [Fe/H]
resulting from the standard model in Fig. \ref{ofe2} (solid line).  We
have then run the same model without SNeIa (dot-short-dashed line) and
without SNeIa and without galactic winds (long-dashed line).  We have
also considered a model without SNeIa in which the ejection efficiency
of $\alpha$-elements is enhanced (selective winds) (short-dashed
line).  Similarly to what we have seen in Fig. \ref{ofe}, if we
neglect SNeIa, there is a negligible variation of the [O/Fe] ratio as
a function of [Fe/H] (dot-short-dashed and long-dashed lines) and a
knee can be attained only in the presence of $\alpha$-enhanced winds,
namely winds in which O is ejected out of the galaxy more efficiently
than Fe ($\alpha_O=2 \alpha_{Fe}$; short-dashed line).  The turnoff
point of this curve corresponds to the moment in which the galaxy
experiences the galactic wind.  This plot shows that only a model
which takes properly into account SNe Ia and galactic winds is able to
produce the knee in [$\alpha$/Fe] observed in several dwarf spheroidal
galaxies.

\section{Discussion and conclusions}
\label{discussion}

In this paper we have studied general properties of simple chemical
evolution models of galaxies with infall and differential winds,
namely with a selective loss of metals from the parent galaxy.  We
have seen that, assuming gas flows proportional to the SFR, the
equations of the evolution of gas mass, total mass and metallicity Z
of gas can be solved analytically, leading to a simple expression of
the variation of Z as a function of the gas mass fraction $\mu$ and of
the ejection efficiency of metals $\alpha$ (which is defined as the
ratio between the metallicity of the galactic wind and the metallicity
of the ISM).  As expected, the resulting metallicity of the galaxy
turns out to be systematically lower than the metallicity attained by
models with non-differential winds.  In particular, the metallicity of
models with differential winds tends asymptotically (for low values of
$\mu$, namely during the late evolution of the galaxy) to a constant
value which decreases almost linearly with $\alpha$.  We have
checked that, for exponential infall laws and star formation rates
proportional to the gas mass (linear Schmidt law) a constancy between
the star formation rate and the infall rate arises quite naturally
during the late evolution of the galaxies, provided that the star
formation timescale does not differ significantly from the infall
timescale.  In the first few Gyr the ratio between infall rate and
star formation rate can vary quite significantly but, in spite of
that, the resulting evolution of the metallicity as a function of time
does not change appreciably compared to the case with linear flows.

We have also analyzed models in which the metallicity of the infalling
gas varies with time.  A realistic case of infalling gas with
increasing metallicity can be envisaged in the interaction galactic
winds/ICM.  Indeed, this interaction must certainly lead to some
pollution of the gas external to the galaxy, therefore to a constant
increase of the ICM metallicity, in particular when differential winds
are considered.  In order to take it into account, we have made two
extreme assumptions:

\begin{itemize}

\item the metals carried out of the galaxy instantaneously and 
uniformly mix with the surrounding ICM ({\it instantaneous ICM
mixing}).

\item The metallicity of the infalling gas is set to be equal to 
the metallicity of the outflow, representing a physical situation in
which the very same gas carried out of the galaxy by outflows (or a
fraction of it) falls back to the galaxy due to the pull of its
gravitational potential ({\it galactic fountain}).

\end{itemize}
\noindent
Real physical situations should lie between these two extreme
assumptions since the distribution of metals in the ICM is expected to
be patchy (Domainko et al. \cite{dom06}; Sanders \& Fabian
\cite{sf06}; Tornatore et al. \cite{tor07}).  The metallicities
predicted by instantaneous ICM mixing models are of course larger than
models in which the infalling gas has a pristine chemical composition.
The magnitude of this difference depends on the relative amount of ISM
and ICM masses.  Under normal circumstances, in which the ICM mass is
larger than the ISM one, the difference in the final metallicity is
only $\sim$ 0.1 - 0.2 dex.  A larger ISM metallicity is obtained by
galactic fountains models.  This result due to the fact that galactic
fountains maximize the chemical enrichment of the infalling gas.  This
metallicity tends however to the one attained by the model with $Z_A =
0$ if very large metal ejection efficiencies ($\alpha \gg 1$) are
assumed.

We have then studied the metallicity ratios produced by differential
wind models.  We have seen that the models with variable infall
metallicities (both instantaneous ICM mixing and galactic fountain)
simply predict that abundance ratios between chemical elements are
given by the ratio of their respective stellar yields (if these do not
change with metallicity).  Assuming a fixed infall metallicity $Z_A
\neq 0$ this ratio also depends on the value of $Z_A$ and on the
infall parameter $\Lambda$, but nevertheless it does not change during
the evolution of the galaxy.  However, we stress here that simple
models can be properly applied only to SNeII products, whose release
is very prompt.  Iron is for instance mostly produced by SNeIa, whose
explosion timescales range from a few tens of Myr to several Gyr,
depending on the star formation history of the considered galaxy
(Matteucci \& Recchi \cite{mr01}), therefore only the fraction of iron
produced by SNeII ($\sim$ 1/3 of the total) can be treated by the
simple models.  A constant O/Fe ratio, as predicted by simple models
is therefore expected because, in this case, both chemical elements
are produced by massive stars, on similar timescales and they should
therefore have a similar evolution.  However, O/Fe ratios are not
constant in galaxies and larger ratios are observed in older, more
metal-poor stars.  The most natural explanation of this trend is that
the delayed production of iron by SNeIa reduces the initially high
O/Fe ratio of the oldest stars ({\it time-delay model}; Matteucci \&
Greggio \cite{mg86}).

Simple models can produce variable abundance ratios either if we
assume different ejection efficiencies for different chemical species
(selective winds) or an IMF variable in time.  Both assumptions should
be quite ``ad hoc''.  We have tested the case of selective winds
for the O/Fe ratio showing that, if the O ejection efficiency is
significantly larger than the Fe one, O/Fe decreases with time (and
with Fe/H).  We have confirmed this result also by means of detailed
numerical simulations in which IRA is relaxed showing that, neglecting
SNeIa, the only way to get a knee in the [O/Fe] vs. [Fe/H] relation is
by assuming $\alpha$-enhanced galactic winds.  In a model suitable for
dwarf spheroidal galaxies, in which the outflow rate exceeds the
infall rate, the O/Fe is almost constant up to [Fe/H] $\sim$ -1.5,
then it bends down.  This behavior is consistent with the observation
of many dwarf spheroidal galaxies of the Local Group (Venn et al.
\cite{venn04} and references therein).  If instead iron is ejected
more efficiently than oxygen, the opposite happens, with O/Fe ratios
increasing with time.  On the other hand, an iron ejection efficiency
larger than the oxygen seems plausible, as it has been shown by
chemodynamical simulations of dwarf galaxies experiencing violent
bursts of star formation (Recchi et al.  \cite{rmd01}; \cite{rec02}).
However, also these models predict O/Fe ratios decreasing with time
(and with [Fe/H]) because SNeIa were not neglected and their
contribution is more important than selective winds in determining the
evolution of the O/Fe ratio.  Moreover, Recchi et al.'s model found
larger iron ejection efficiencies precisely because of the delayed
explosion of SNeIa.  These SNe explode in a medium already heated and
diluted by the previous activity of SNeII, therefore SNeIa products
can be very easily channeled along the galactic funnel.  If we neglect
SNeIa, O and Fe are produced by the same stars and they should
therefore share the same fate.

Our main conclusions can be summarized as follows:

\begin{itemize}

\item
 
  The differential winds have the effect of reducing the global
  metallicity of a galaxy.  The magnitude of this reduction increases
  with the ejection efficiency of metals.

\item

  This remains true also if we allow a variable metallicity of the
  infalling gas.  Larger galactic metallicities are obtained if we
  assume that the infall has the metallicity of the outflowing gas
  (galactic fountain model).

\item 

  Simple models with constant (i.e. non metallicity-dependent) yields
  predict constant abundance ratios, even in the presence of
  differential winds or variable infall metallicities.  The same
  occurs for detailed numerical models.

\item

  One way to get variable abundance ratios in simple models is by
  assuming selective winds, namely differential winds with different
  ejection efficiencies for each chemical element.  However, simple
  models apply only to chemical elements promptly released on short
  timescales, therefore large element-to-element variations of the
  ejection efficiencies are not very likely.  Abundance ratios
  predicted by simple models of chemical evolution are then likely to
  remain constant during the evolution of the galaxy. 

\end{itemize}

\begin{acknowledgements}

We thank the anonymous referee whose comments and criticisms have
improved the quality of the paper. We thank F. Calura, G. Cescutti and
A. Palestini for important support and stimulating discussions.
G.A.L.  acknowledges financial support from the Brazilian agency
FAPESP (proj. 06/57824-1).
  
\end{acknowledgements}

\section{Appendix}
\label{sec:app}

Here we give the explicit value of some constants we have used in
Sect. \ref{sec:limits}.

\begin{equation}
E_1 = (\lambda + 1) (1 - R) S t_i - 1.
\end{equation}

\begin{equation}
E_2 = \lambda (\alpha - 1) (1 - R) S t_i.
\end{equation}

\begin{equation}
K_1 = \gamma E_1 - 1.
\end{equation}

\begin{equation}
K_2 = {K_1 \over {\lambda (\alpha - 1)}}.
\end{equation}

\begin{equation}
K_3 = {{(1 - R) S t_i} \over {(1 - R) (1 + \alpha \lambda) S t_i - 1}}.
\end{equation}

\end{document}